\documentclass[prl, twocolumn]{revtex4}
\usepackage{amssymb}
\usepackage{amsmath}
\usepackage{tabularx}
\usepackage{graphicx}
\usepackage{color}
\usepackage{verbatim}
\usepackage{epstopdf}

\begin{document}

\title{Revealing the origin of super-Efimov states in the hyperspherical formalism}
\author{Chao Gao}
\affiliation{Institute for Advanced Study, Tsinghua University, Beijing 100084, China}
\author{Jia Wang}
\affiliation{Department of Physics, University of Connecticut, Storrs, Connecticut 06269, USA}
\author{Zhenhua Yu}
\email{huazhenyu2000@gmail.com}
\affiliation{Institute for Advanced Study, Tsinghua University, Beijing, 100084, China}

\date{\today}
\begin{abstract}
Super-Efimov states are a new kind of universal three-body bound states predicted for three identical fermions with $p$-wave resonant interactions in two dimensions by a recent field-theoretic calculation [Phys.~Rev.~Lett.~\textbf{110}, 235301 (2013)]. The binding energies of these states obey a dramatic double exponential scaling $E_n=E_*\exp(-2 e^{\pi n/s_0+\theta})$ with universal scaling $s_0=4/3$ and three-body parameters $E_*$ and $\theta$. We use the hyperspherical formalism and show that the super-Efimov states originate from an emergent effective potential $-1/4\rho^2-(s_0^2+1/4)/\rho^2\ln^2\left(\rho\right)$ at large hyperradius $\rho$. Moreover, for pairwise interparticle potentials with van der Waals tails, our numerical calculation indicates that the three-body parameters $E_*$ and $\theta$ are also universal and the ground super-Efimov state shall cross the threshold when the $2$D $p$-wave scattering area is about $-42.0\, l_\text{vdW}^2$ with $l_\text{vdW}$ the van der Waals length. \end{abstract}
\maketitle

A landmark result of few-body physics is the Efimov bound states predicted theoretically in 1970 for three-body systems with $s$-wave resonant interactions in three dimensions \cite{efimov}. The binding energy of the $n$th Efimov state scales as $E_n\sim \tilde E_* e^{-2\pi n/\tilde s_0}$ with $\tilde s_0$ a universal number and $\tilde E_*$ the three-body parameter \cite{efimov, braaten, petrov}. This peculiar scaling is given rise to by an emergent effective potential of the form $-(\tilde s_0^2+1/4)/\rho^2$ in the hyperspherical formalism of the three-body problem at large hyperradius $\rho$.
Only recently, extreme experimental controllability and versatility of ultra-cold atomic gases \cite{bloch, molmer, dalibard} provides a unique opportunity to detect evidences of the Efimov states for the very first time in atomic systems.
Experimentalists succeeded in realizing $s$-wave resonant interactions in ultra-cold atomic gases by the technique of Feshbach resonance \cite{Chin}, and revealed the Efimov physics through measuring atom loss rate due to three-body recombinations \cite{grimm_cesium, Jochim_3b, hulet, Gross1, OHara1, OHara2, Jin}, atom-dimer inelastic collisions \cite{grimm_ad, jochim_ad} and radio-frequency spectroscopy \cite{jochim_radio, ueda_radio}. Further studies showed that even the three-body parameter $\tilde E_*$ which determines the absolute energy levels of the Efimov states has a universal behavior for different atomic species \cite
{Berninger,Chin_a,Wang1, Wang2,Schmidt,Naidon}.

The quest for universal physics at resonances beyond the paradigm of the Efimov states brought about a recent quantum field theory calculation predicting that universal bound states exist for three identical fermions with  $p$-wave resonant interactions in two dimensions \cite{nishida_super}. These new states have angular momentum $\ell=\pm1$ and are called ``super-Efimov" due to the fascinating scaling of their binding energies $E_n=E_*\exp(-2e^{n\pi/s_0+\theta})$ with $s_0=4/3$ a universal number, and $E_*$ and $\theta$ the three-body parameters. While the prediction of the super-Efimov states agrees with a recently proved theorem \cite{frankfurt}, understanding the origin of such universal states requests further investigation.

In this work, we use the hyperspherical formalism to study three identical fermions with $p$-wave resonant interactions in two dimensions. In the angular momentum $\ell=\pm1$ channel, we show that the super-Efimov states are due to an emergent effective potential $U_{\rm eff}\sim-1/4\rho^2-(s_0^2+1/4)/\rho^2\ln^2(\rho)$ in the large hyperradius $\rho$ limit. We extract $s_0$ from $U_{\rm eff}$ calculated numerically at the first three $p$-wave resonances of three different kinds of model potentials; the extracted values of $s_0$ agree well with $4/3$ as predicted by the field theory \cite{nishida_super}. For pairwise interparticle potentials with a van der Waals tail, the numerically obtained binding energies of the lowest two super-Efimov states indicate that the three-body parameters $E_*$ and $\theta$ are also universal; the ground super-Efimov state is predicted to emerge at the threshold when the $2$D scattering area is about $-42.0\,l_\text{vdW}^2$ with $l_\text{vdW}$ the van de Waals length. 

\emph{Hyperspherical formalism.---}
We consider three identical fermions with coordinates $\mathbf r_1$, $\mathbf r_2$ and $\mathbf r_3$ interacting pairwisely through a central potential $V(r)$ of finite range $r_0$ in two dimensions. The potential is fine tuned such that it is at a $p$-wave resonance. We introduce the Jacobi coordinates $\mathbf x_i=\mathbf r_j-\mathbf r_k$ and $\mathbf y_i=2[\mathbf r_i-(\mathbf r_j+\mathbf r_k)/2]/\sqrt{3}$, where $\{i,j,k\}$ takes the values of $\{1,2,3\}$ cyclically. The hyperspherical radius is given by $\rho=\sqrt{\mathbf x_i^2+\mathbf y_i^2}$, and the corresponding hyperspherical angles $\Omega_i=\{\alpha_i, \theta_{\mathbf x_i},\theta_{\mathbf y_i}\}$ with $\alpha_i= \tan^{-1}(x_i/y_i)$, 
$\theta_{\mathbf x_i}$ ($\theta_{\mathbf y_i}$) the polar angle of $\mathbf x_i$ ($\mathbf y_i$) \cite{nielsen}.
After separating out the center of mass part, we expand the wave-function of the system in terms of any set of hyperangles $\Omega_i$ as
\begin{align}
\Psi=\sum_\mu\rho^{-3/2}f_\mu(\rho)\Phi_\mu(\rho,\Omega_i).
\end{align}
The angular part $\Phi_\mu(\rho,\Omega_i)$ is required to satisfy the eigenequation
\begin{align}
\left[\hat\Lambda^2+m\rho^2\sum_{j=1}^3 V(\rho\sin\alpha_j)\right]\Phi_\mu(\rho,\Omega_i)=\lambda_\mu(\rho)\Phi_\mu(\rho,\Omega_i),\label{aeq}
\end{align}
with $m$ the mass of each fermion. Here, the total angular momentum operator is given by \cite{nielsen}
\begin{align}
\Lambda^2=-\frac{\partial^2}{\partial\alpha_i^2}-2\cot(2\alpha_i)\frac{\partial}{\partial\alpha_i}+\frac{L^2_{\mathbf x_i}}{\sin^2\alpha_i}+\frac{L^2_{\mathbf y_i}}{\cos^2\alpha_i}.
\end{align}
Hereafter, we use units such that $\hbar=m=1$. 
Consequently, the hyperradial part satisfies the coupled equations of eigenenergy $E$ as \cite{nielsen}
\begin{align}
&\left[-\frac {d^2}{d\rho^2}-\frac1{4\rho^2}+U_\mu(\rho)-Q_{\mu\mu}-E\right]f_\mu(\rho)\nonumber\\
&=\sum_{\nu(\neq \mu)}\left[2P_{\mu\nu}\frac d{d\rho}+Q_{\mu\nu}\right]f_{\nu}(\rho), \label{feq}
\end{align}
with $U_\mu(\rho)=[\lambda_\mu(\rho)+1]/{\rho^2}$.
The couplings $P_{\mu\nu}=\langle \Phi_\mu|\partial_\rho|\Phi_{\nu}\rangle$ and $Q_{\mu\nu}=\langle \Phi_\mu|\partial^2_\rho|\Phi_{\nu}\rangle$, with $\langle\dots\rangle$ standing for the integration over the hyperangles, are expected to be negligible for $\mu\neq \nu$ in the large $\rho$ limit \cite{nielsen}; 
as Eq.~(\ref{feq}) becomes decoupled, the three-body problem is reduced to a one dimensional equation, 
and the eigenstates with $E\to0^-$ shall be governed by the effective potential 
\begin{equation}
U_{\rm eff}(\rho)\equiv -\frac{1}{4\rho^2}+U_0-Q_{00}
\label{Ueff}
\end{equation}
of the shallowest attractive channel $\mu=0$ at large hyperradius.

We focus on the states with total angular momentum $|\ell|=|\ell_{\mathbf x_i}+\ell_{\mathbf y_i}|=1$ for which the super-Efimov states were predicted \cite{nishida_super}. We solve the Faddeev equations derived from Eq.~(\ref{aeq}) in the regime $r_0/\rho\ll1$ \cite{nielsen}, and find for the shallowest attractive channel \begin{align}
\lambda_0(\rho)+1=-\frac Y{\ln(\rho/r_0)}+O\left(\frac1{\ln^2(\rho/r_0)}\right)\label{lambda0},
\end{align}
where the dimensionless parameter $Y$ is given by
\begin{align}
Y=&-1-\frac{\int^\infty_0 dr r^2 V(r) u_0^2(r)}{\lim_{r\to\infty}[r u_0(r)^2]}\label{y}
\end{align}
with $u_0\left({r}\right)$ the zero energy $p$-wave two-body reduced radial wave function satisfying 
\begin{equation}
\left[-\partial^2_r+\frac3{4r^2}+V(r)\right]u_0(r)=0.
\end{equation}
The radial effective potential $V(r)+3/4r^2$ has a centrifugal barrier. An alternative expression is \cite{zinner, castin}
\begin{align}
Y=&\frac{\int^\infty_0 dr r  \left[\partial_r( u_0(r)\sqrt{r})\right]^2}
{\lim_{r\to\infty}[\sqrt{r} u_0(r)]^2}\label{ya},
\end{align}
which shows $Y$ positive definite. Note that a similar logarithmic structure also appears in the scattering $T$-matrix in two dimensions \cite{jesper}.

\emph{Effective potential.---}
In the regime $r_0/\rho\ll1$, if $Q_{00}$ can be neglected, $U_{\rm eff}+1/4\rho^2\sim-Y/\rho^2\ln(\rho/r_0)$ would give rise to shallow bound states whose energies $E_n$ scale as $\ln|E_n|\sim -(n\pi)^2/2Y$. Surprisingly Ref.~\cite{zinner} argued that $Q_{00}\sim-Y/\rho^2\ln(\rho/r_0)$; the leading orders of $U_0$ and $Q_{00}$ shall cancel. This cancellation would result in $U_{\rm eff}+1/4\rho^2=U_0-Q_{00}\sim1/\rho^2\ln^2(\rho/r_0)$ in which case super-Efimov states become possible.

The involved hyperangle integral of $Q_{00}$ seems to preclude evaluating it analytically to order $1/\rho^2\ln^2(\rho/r_0)$. Hence we obtain $U_{\rm eff}$ by calculating $U_0$ and $Q_{00}$ numerically with three kinds of model potentials: the Lennard-Jones potential (LJ) ${V_{\rm LJ}}\left( r \right) =  - {V_0}\left[ {{{\left( {{r_0}/r} \right)}^6} - {\eta ^6}{{\left( {{r_0}/r} \right)}^{12}}} \right]$, the Gaussian potential (GS) ${V_{\rm GS}}\left( r \right) =  - {V_0}\exp \left[ { - {{\left( {r/r_0} \right)}^2}} \right]$, and the P\"oschl-Teller potential (PT) ${V_{\rm PT}}\left( r \right) =  - {V_0}{\rm sech}^2\left( {r/{r_0}} \right)$. The model potentials are all tuned at a $p$-wave resonance.
We solve Eq.~(\ref{aeq}) numerically by using the modified Smith-Whitten coordinates, which have been successfully applied to three-body systems in both three dimensions \cite{Johnson,Lepetit,Lin,Suno,Wang_Dincao} and two dimensions \cite{Dincao1,Dincao2}. The details of constructing the Smith-Whitten coordinates and the corresponding hyperspherical representation can be found  in Refs.~\cite{Dincao1} and \cite{Wang_Wang}.

\begin{figure}
\includegraphics[width=8cm]{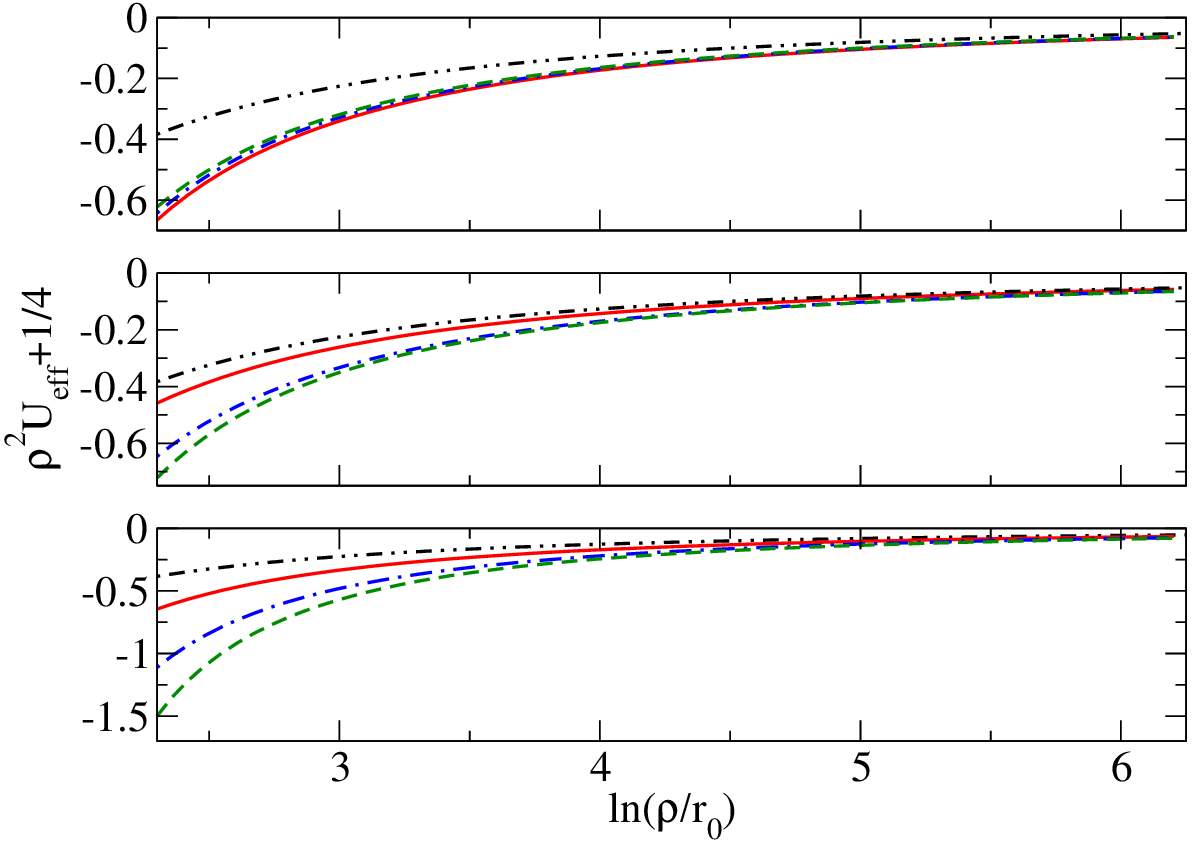}
\caption
{Numerical results for the effective potential $U_{\rm eff}$ defined in Eq.~(\ref{Ueff}) for three different two-body model potentials from top to bottom: Lennard-Jones (LJ), Gaussian (GS), P\"oschl-Teller (PT). The red solid lines are for the first $p$-wave resonances of the three potentials, and the blue dash-dotted ones for the second, and the green dashed ones for the third. The black dash-dot-dotted lines are $\rho^2U_{\rm eff}+1/4=-[(4/3)^2+1/4]/\ln^2(\rho/r_0)$.}
\label{numerical}
\end{figure}

Figure (\ref{numerical}) shows the resultant numerical results of $U_{\rm eff}$ at the first three $p$-wave resonances of the three model potentials, which all converge to a universal form $-1/4\rho^2-[(4/3)^2+1/4]/\rho^2\ln^2(\rho/r_0)$ when $\rho/r_0$ is large.
We fit the data of $\rho^2U_{\rm eff}+1/4$ by the series $-\sum_{n=2}^4 c_n \ln^{-n}(\rho/r_0)$ in the range $\rho/r_0\in[30,500]$. We define $s_0^2\equiv c_2-1/4$. Table~(\ref{fit}) shows that all fitted values of $s_0$ agree with $4/3$ within $\sim4\%$.
Likewise we fit the data for $\rho^2U_0$ and $\rho^2Q_{00}$ separately by $-\sum_{n=1}^3 c_n \ln^{-n}(\rho/r_0)$ in the same range.
As shown in Tab.~(\ref{fit}), fitted $c_1$ of both $U_0$ and $Q_{00}$ and $Y$ calculated by the analytic result Eq.~(\ref{y}) show good agreement within $\sim6\%$, the difference between which nevertheless quantifies the overall error of our numerical data and the fitting scheme.

\begin{center}
\begin{table}
\caption{The parameter $Y$ calculated from Eq.~(\ref{y}) and the fitted parameters to the numerical results for different model potentials from the first to the third $p$-wave resonance.}\label{fit}
\begin{ruledtabular}
\begin{tabular}{ c|c c c c }
Resonance & $Y$ & $c_1$ of $U_0$ & $c_1$ of $Q_{00}$ & $s_0$ of $U_{\rm eff}$ \\
\hline
LJ $1$st & $1.068$ &1.063 & 1.071  & 1.339\\
\hline
LJ $2$nd & $1.939$ & 1.979 & 1.960  & 1.348\\
\hline
LJ $3$rd & $2.393$ & 2.519 & 2.452 & 1.381 \\
\hline
GS $1$st & $0.484$ & 0.475 & 0.484 & 1.341\\
\hline
GS $2$nd & $1.636$ & 1.654 & 1.641 & 1.355\\
\hline
GS $3$rd & $2.781$ & 2.949  & 2.872 & 1.393 \\
\hline
PT $1$st &  $0.437$ & 0.431& 0.437 & 1.350\\
\hline
PT $2$nd &  $1.209$ & 1.209 &1.209 & 1.349\\
\hline
PT $3$rd & $1.880$  & 1.928 & 1.885 & 1.367 
\end{tabular}
\end{ruledtabular}
\end{table}
\end{center}

Our calculation indicates that when $\rho/r_0$ is large, the three-body system is subject to an emergent effective potential
\begin{align}
U_{\rm eff}(\rho)=-\frac1{4\rho^2}-\frac{s_0^2+1/4}{\rho^2\ln^2(\rho/r_0)}.\label{uadiab}
\end{align}
Given such a potential, one can use the WKB approximation (or other methods) to show that the binding energies of shallow bound states have the super-Efimov form $E_n=E_*\exp(-2 e^{\pi n/s_0+\theta})$. Our numerical results of $s_0$ agrees well with the universal scaling factor $4/3$ predicted by Ref.~\cite{nishida_super}. Thus we show that the universal super-Efimov states originate from the universal effective potential Eq.~(\ref{uadiab}).

The above conclusion is based on the adiabatic approximation by neglecting inter-channel couplings [cf.~Eq.~(\ref{Ueff})]. For the Lennard-Jones, Gaussian, and P\"oschl-Teller two-body model potentials, we find numerically that the inter-couplings between the super-Efimov channel $\mu=0$ and other channels $\nu \neq 0$ have the asymptotic behaviors $P_{0\nu}\sim1/\rho\ln^2(\rho)$ and $Q_{0\nu}\sim1/\rho^2\ln^2(\rho)$ when $\rho$ is large. 
The effects of these nonzero inter-channel couplings on the super-Efimov states can be evaluated perturbatively in the following way. First we solve Eq.~(\ref{feq}) at zero order by neglecting all the inter-channel couplings. The $\mu=0$ channel would produce the super-Efimov bound state solutions $f^{(0)}_0(\rho)$ with negative eigenenergies $E$ while apart from any accidental coincidences, in any other channels $\nu\neq0$ there is only a trivial solution $f^{(0)}_\nu(\rho)=0$ for the same energies $E$. Next we substitute $f^{(0)}_0(\rho)$ into Eq.~(\ref{feq}) and solve $f^{(1)}_\nu(\rho)$ for $\nu\neq0$ to the first order of the inter-channel couplings. In the regime $r_0\ll\rho\ll 1/|E|$,  $f^{(0)}_0(\rho)\sim\sqrt{\rho\ln(\rho/r_0)}\cos\{s_0\ln[\ln(\rho/r_0)]+\varphi\}$ with $\varphi$ a phase shift \cite{zinner}, which indicates $f^{(1)}_\nu(\rho)\sim f^{(0)}_0(\rho)/\ln^2(\rho/r_0)$. Thus in Eq.~(\ref{feq}) the off diagonal terms are expected to be $[2P_{0\nu}(d/d\rho)+Q_{0\nu}]f^{(1)}_{\nu}(\rho)\sim f^{(0)}_0/\rho^2\ln^4(\rho/r_0)$, negligible compared with the diagonal term $U_{\rm eff}(\rho)f^{(0)}_0(\rho)$; the adiabatic approximation is justified in the regime $\rho\to\infty$.

\emph{Three-body parameters.---}
In the case of Efimov states, the three-body parameter $\tilde E_*$ is originally believed to be \emph{not universal} and to be determined by short-range interaction details \cite{braaten}. Surprisingly recent experiments of ultra-cold atomic gases found $\tilde E_*$ rather universal (in van der Waals units) \cite{Berninger}. Subsequent theoretical calculations \cite{Chin_a,Wang1,Wang2,Schmidt,Naidon} inspired by this new discovery soon confirmed that when the long range tail of the two-body interaction is dominated by the van der Waals form $V\left({r}\right)\rightarrow-C_6/r^6$, $\tilde E_*$ is universally determined by the van der Waals length $l_{\rm vdW}\equiv C_6^{1/4}/2$ or equivalently the van der Waals energy $E_{\rm vdW}\equiv-1/{l_{\rm vdW}^2}$. This universality of $\tilde E_*$ is attributed to the suppressed probability of finding two particles at short distances where $V(r)$ shows a deep attractive well \cite{Wang1}.
It is natural to ask the question: whether the three-body parameters for super-Efimov states $E_*$ and $\theta$ are also universal, if the two-body interaction has the long-range tail $-C_6/r^6$?

\begin{figure}
\includegraphics[width=8cm]{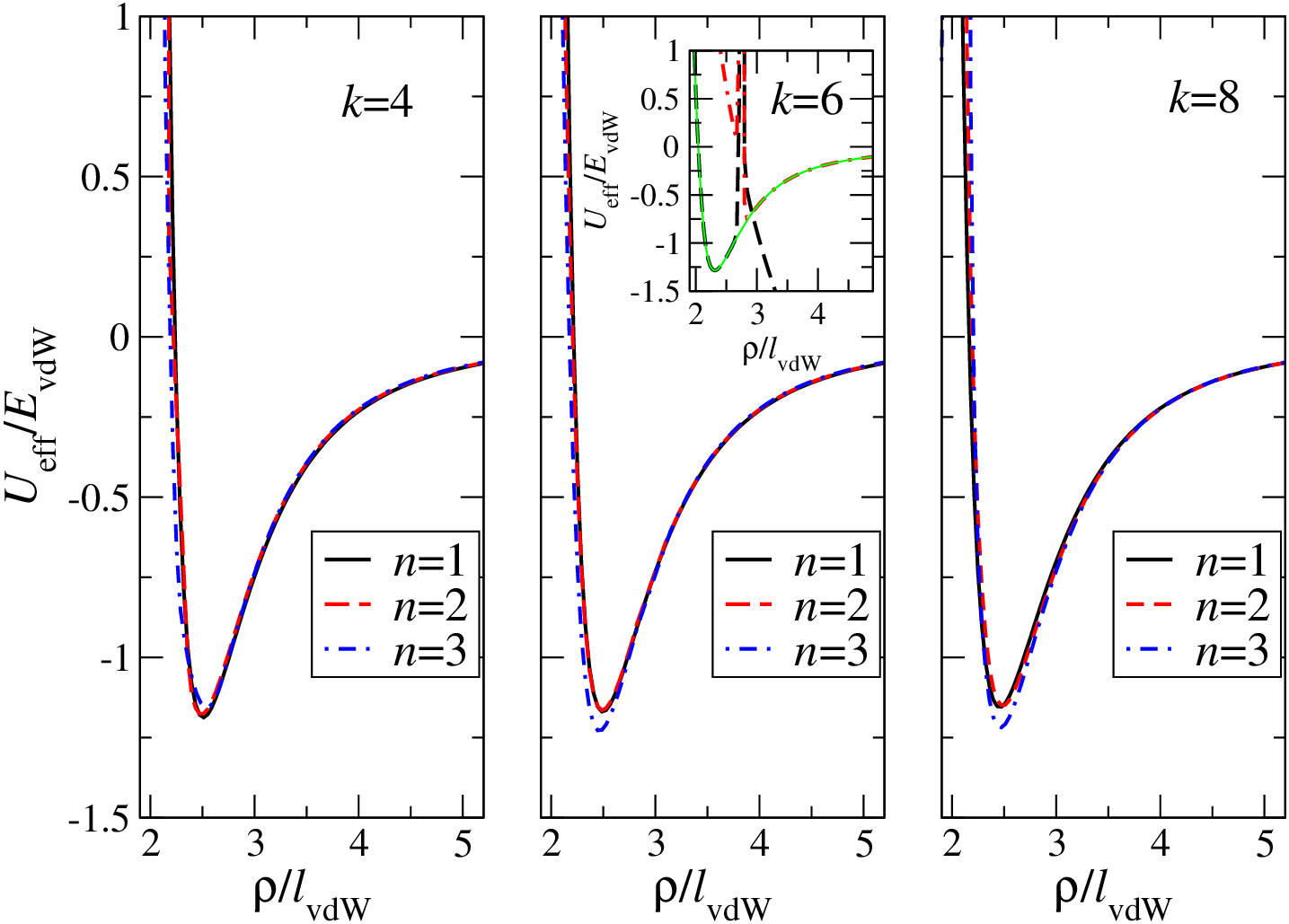}
\caption{Universal effective potential $U_{\rm eff}$ for different two-body model potential $V_k^n$, with sharp avoid crossings manually diabatized in some cases to improve visualization. 
An example of the manual diabatization is shown in the inset for the model potential $V_6^2$. The sharp feature arising from an accidental crossing between two channels represented by the red dash-doted and the black dashed curves is manually eliminated to give the smooth green solid curve.
}\label{Ueffshortrange}
\end{figure}

We use two-body model potentials $V_k^n\left({r}\right)=-C_6/r^6\left[{1-(\beta_n/r)^k}\right]$ to study the three-body parameters numerically. The short-range parameter $\beta_n$ is tuned such that there are $n$ $p$-wave two-body bound states including the shallowest one at threshold. These two-body model potentials have the same long-range van der Waals tail, but very different short-range interactions determined by $\beta_n$ and $k$. The first evidence of universality is the effective potential $U_{\rm eff}$ at short range as shown in Fig.~(\ref{Ueffshortrange}), where a universal repulsive core rises up at about $\rho \approx 2.2 l_{\rm vdW}$; it seems that the short range details of these different two-body model potentials have little effect on those of the three-body effective potential $U_{\rm eff}$. In plotting $U_{\rm eff}$ in Fig.~(\ref{Ueffshortrange}), we have 
manually diabatized the curves to improve visualization. 
One example is shown in the inset of Fig.~(\ref{Ueffshortrange}), where a sharp feature arising from an accidental crossing between the super-Efimov channel and another channel is manually eliminated. This kind of sharp features of $U_{\rm eff}$ at small $\rho$ shall not be important for understanding low-energy three-body observables.

Applying the numerical treatment similar to Ref.~\cite{Wang_Dincao}, we calculate the three-body super-Efimov ground state energies $E_g$ for different $V_k^n\left({r}\right)$.  
When the model potentials $V^n_k(r)$ can support only one two-body bound state at threshold ($n=1$), the super-Efimov channel is the lowest three-body channel; the super-Efimov states are \emph{true} bound states and we obtain their eigenenergies by diagonalizing the Hamiltonian directly. When the model potentials support multiple two-body bound states ($n>1$), deeper three-body channels (atom-dimer channels) exist, and the super-Efimov states become \emph{quasi}-bound states. 
It is known that when there is a quasi-bound state buried in the continua, the scattering amplitude 
shows a Fano resonance due to the interference between the continuum states and the quasi-bound state \cite{Fano}. In this case, we calculate the scattering cross sections for the deeper atom-dimer channels at energies close to those of the super-Efimov states, and locate resonances that can be fitted by a Fano lineshape. The resonance positions are interpreted as the super-Efimov state energies, and the widths of the resonances give rates of the super-Efimov states decaying into atom-dimer states.

\begin{figure}
\includegraphics[width=8cm]{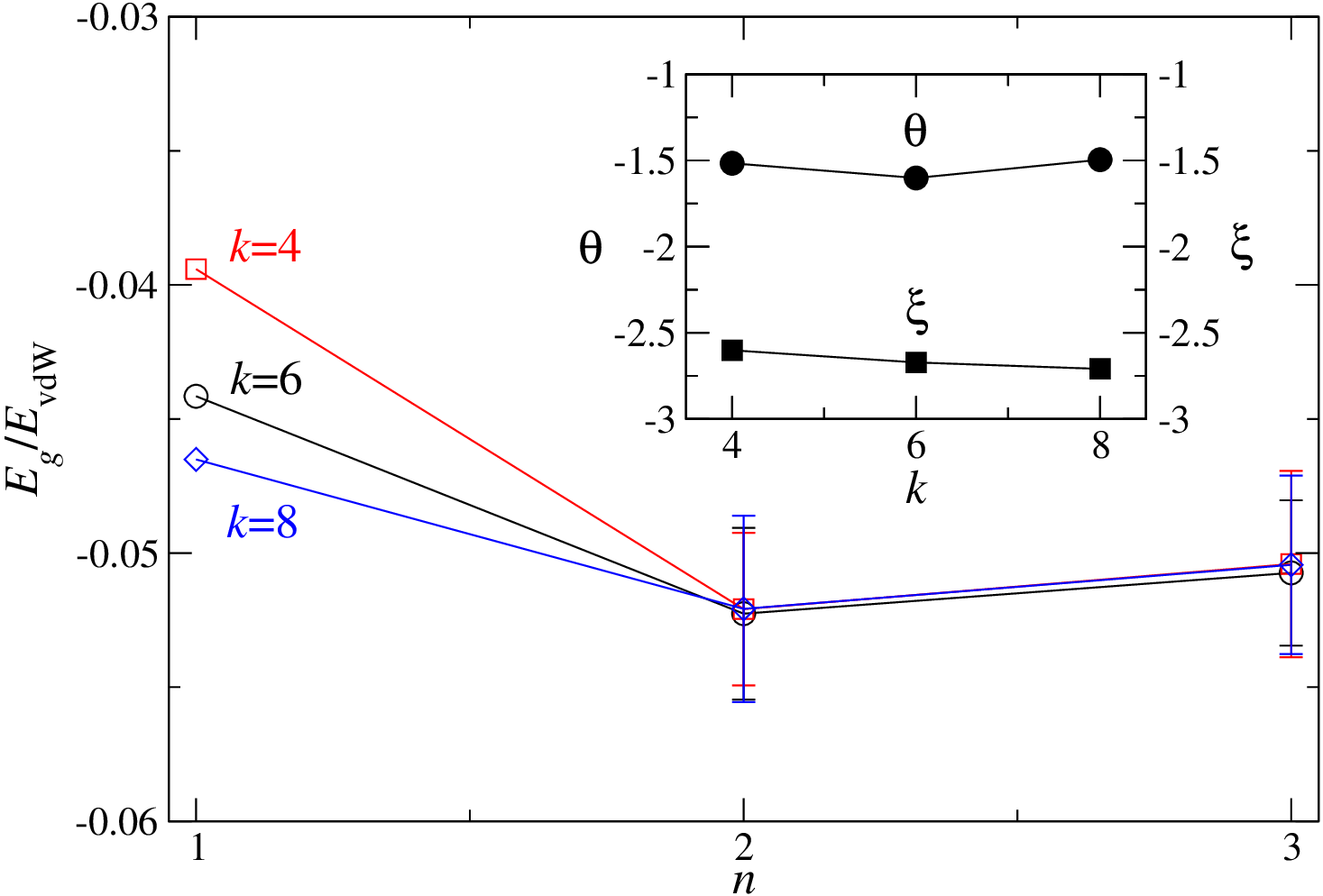}
\caption{Super-Efimov ground state energies $E_g$ for different two-body model potentials $V_k^n$ by full calculations. The error bars at $n=2,3$ are the decay rates of these states into atom-dimer states. The inset shows the three-body parameters $\theta$ and $\xi$ calculated by the adiabatic approximation.}\label{Eg}
\end{figure}

The super-Efimov ground state energies obtained in the way described above are quite universal $E_g/E_{\rm vdW}\approx -0.05$ as shown in Fig.~(\ref{Eg}) and are interestingly close to the universal Efimov ground state energies \cite{Wang1}, while small decay rates are found for $n=2,3$. In addition, we extrapolate $U_{\rm eff}$ to very large distances and calculate the energies $E^{\rm ad}_g$ and $E^{\rm ad}_1$ of both the ground and the first excited super-Efimov states for $V_k^1\left({r}\right)$ within the adiabatic hyperspherical approximation (neglecting $P_{0\nu}$ and $Q_{0\nu}$ for $\nu\neq0$). Note that possible decay rates can not be evaluated within this approximation. 
Table (\ref{para}) shows that while the ground state energies $E^{\rm ad}_g$ have good agreement with the full calculations $E_g$, the first excited state energies $E^{\rm ad}_1$ have extremely small values (of order $10^{-14}E_{\rm vdW}$), implying that a full calculation for the first excited states will be extremely challenging. 
If we express the super-Efimov energies as $E/E_{\rm vdW}=\exp\left[-2\exp\left(4n\pi/3+\theta\right)+\xi\right]$, the three-body parameters $\theta$ and $\xi$ [$\equiv \ln(-E_*/E_{\rm vdW})$] extracted from $E^{\rm ad}_g$ and $E^{\rm ad}_1$ are shown in the inset of Fig.~(\ref{Eg}) to be very universal.

\begin{figure}
\includegraphics[width=8cm]{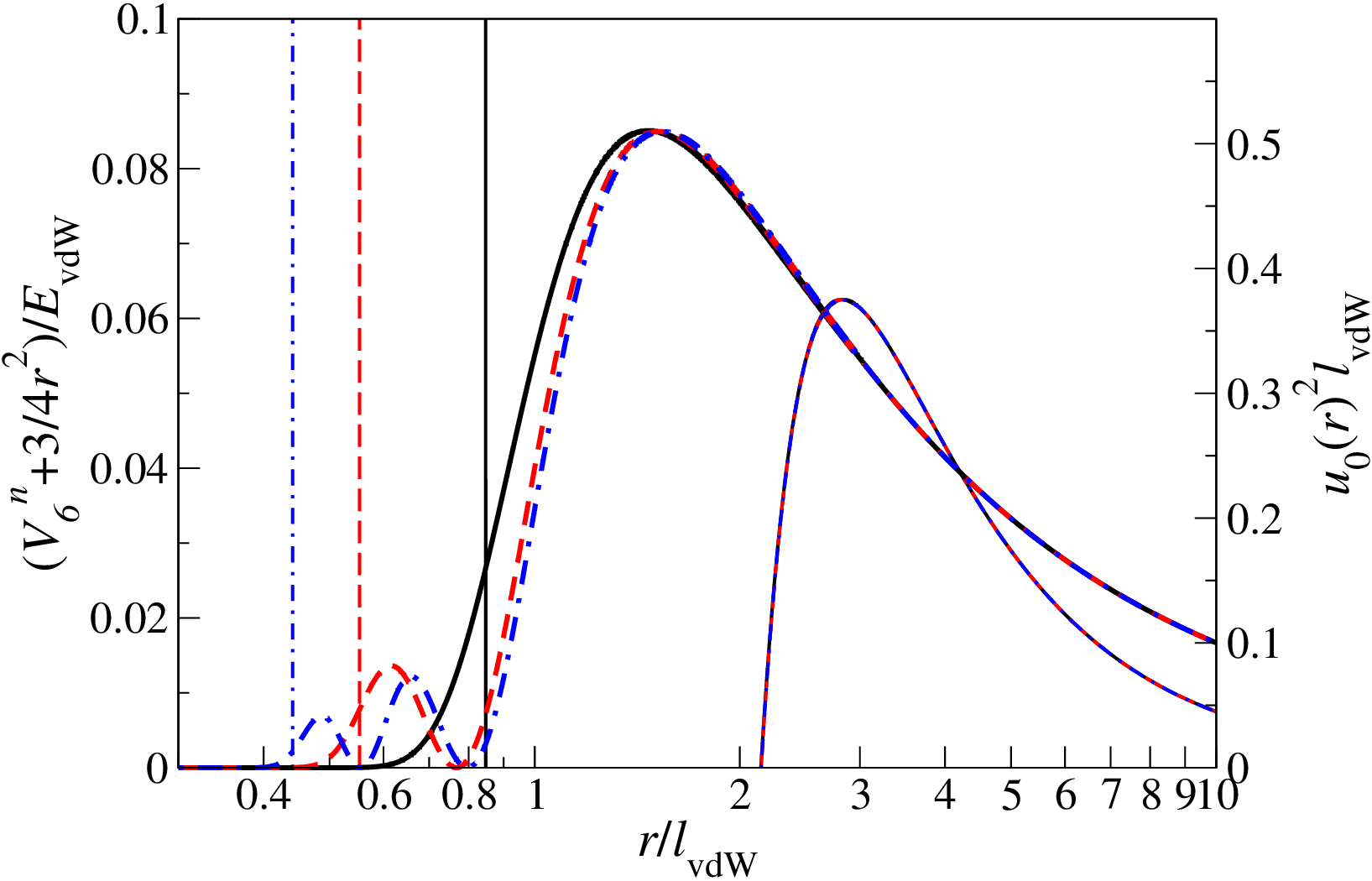}
\caption{Suppression of the zero energy two-body wave function $u_0$ at short distances. The thick curves are the radial probability of zero energy wave functions $u_0^2\left({r}\right)$ for the model potential $V_6^n$ with $n=1,2,3$; the thin curves are the effective radial potential $V_6^n+3/4r^2$. The black solid, red dashed and blue dash-dotted curves correspond to $n=1,2,3$ respectively.}\label{twobody}
\end{figure}

We attribute the universality of $\theta$ and $\xi$ to the mechanism similar as in the Efimov states, i.e., the probability of finding any pair of particles separated by less than $l_{\rm vdW}$ is greatly suppressed, implying that the short distance details of interactions have negligible effects \cite{Wang1}. For example,
We calculate the zero energy reduced two-body wave-function $u_0\left({r}\right)$ for the model potential $V_6^n(r)$ with $n=1,2,3$. As shown in Fig.~(\ref{twobody}), we find that $u^2_0(r)$ has substantial magnitude in the range $1<r/l_{\rm vdW}<2$ inside the centrifugal barrier of the effective radial potential $V_6^n+3/4r^2$, which is due to the resonant tunneling right at $p$-wave resonances. The zero energy reduced two-body wave-function is normalized as $u_0(r)=r^{-1/2}$ when $r\to\infty$. The further deep attractive well of $V_6^n+3/4r^2$ strongly suppresses $u^2_0(r)$ to small values when $r<l_{\rm vdW}$, which can be understood by a semiclassical analysis as in the Efimov case \cite{Wang1}.

\begin{table}
\caption{The super Efimov ground state energies $E_g$ by full calculations and the ground state energies $E_g^{\rm ad}$ and the first excited energies $E_1^{\rm ad}$ calculated by the hyperspherical adiabatic approximation. Here $[n]$ denotes $\times10^n$. $\theta$ and $\xi$ are the two three-body parameters.}\label{para}
\begin{ruledtabular}
\begin{tabular}{c|c c c c c}
k & $E_g/E_{\rm vdW}$ & $E_g^{\rm ad}/E_{\rm vdW}$ & $E_1^{\rm ad}/E_{\rm vdW}$ & $\theta$ & $\xi$ \\
\hline
4 & -3.941[-2] & -4.785[-2] & -1.995[-14] &-1.517 & -2.601\\
6 & -4.415[-2] & -4.429[-2] & -1.232[-14] &-1.502 & -2.672\\
8 & -4.651[-2] & -4.254[-2] & -0.969[-14] &-1.496 & -2.709\\
\end{tabular}
\end{ruledtabular}
\end{table}

\begin{figure}
\includegraphics[width=8cm]
{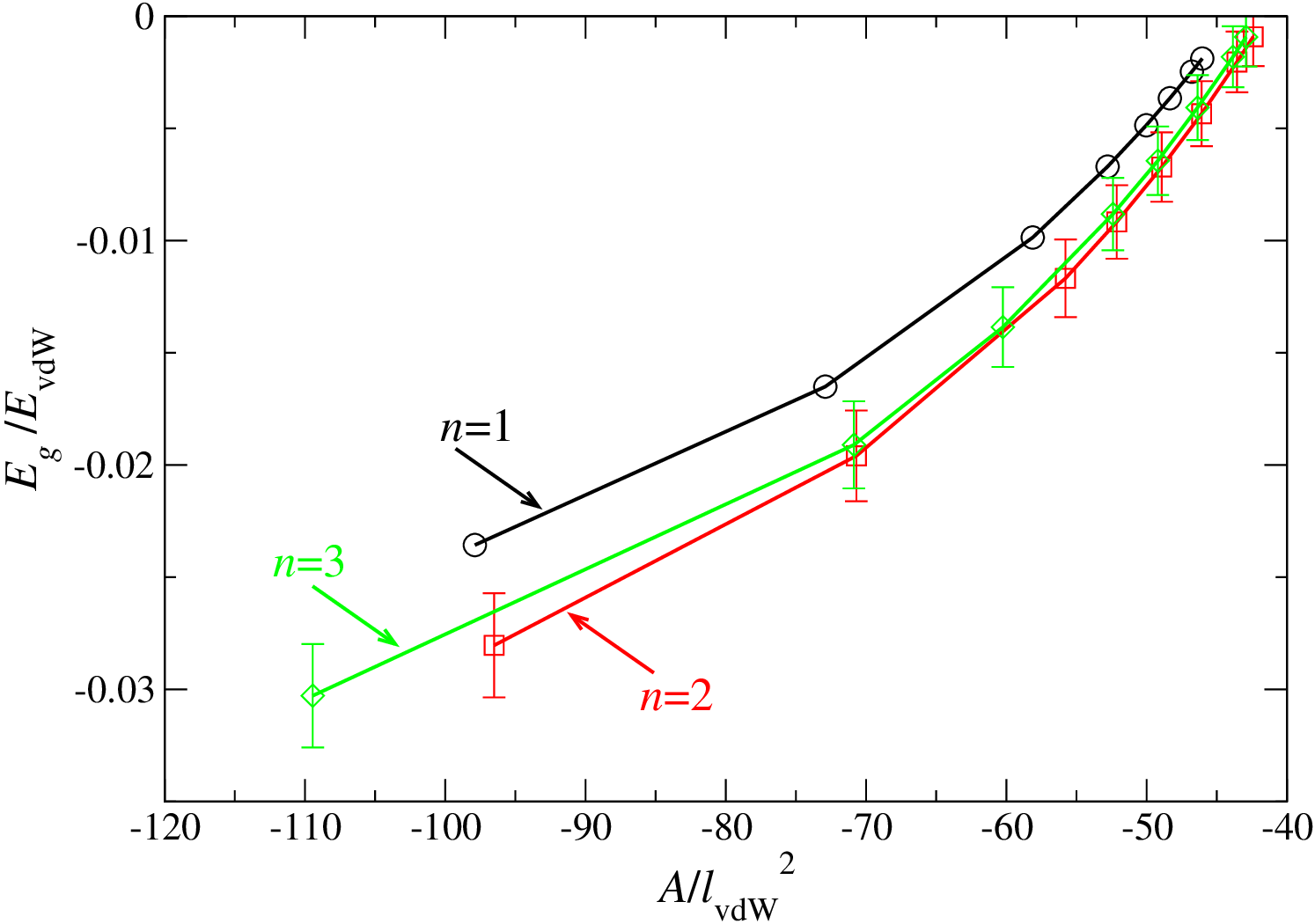}
\caption{The binding energy of the ground super-Efimov trimer state $E_g$ versus the $2$D $p$-wave scattering area $A$ using the Lennard-Jones potential tuned around the $n$th $p$-wave resonance. The error bars of $E_g$ for $n=2,3$ quantifies the finite lifetime of the state due to its decaying into atom-dimer states.}\label{SER}
\end{figure}

\emph{Threshold crossing.}--- In ultra-cold atomic gases, the three-body recombination resonances observed experimentally in the vicinity of Feshbach resonances occur where Efimov state energies cross the three-body continuum threshold, and serve as first evidences of Efimov physics \cite{grimm_cesium, Jochim_3b, hulet, Gross1,OHara1, OHara2, Jin, gross}. Here we tune the depth of the Lennard-Jones two-body model potential around the $n$th $p$-wave resonance, and calculate the ground super-Efimov state energy $E_g$ as a function of $2$D $p$-wave scattering area $A$. [For small scattering wave vector $q$, the $2$D $p$-wave scattering phase shift $\delta(q)$ is given by $\cot\delta(q)=-1/A q^2$.] Figure (\ref{SER}) shows that when $A$ is tuned to large and negative values, $E_g$ becomes shallower and eventually hit the three-body continuum. Extrapolating $E_g$ to the threshold, we find that the crossing point $A^{(-)}_g$ is at 
$-45.9 \,l_\text{vdW}^2$, $-42.1\,l_\text{vdW}^2$, and $-42.0\,l_\text{vdW}^2$ near the $1$st, $2$nd, and $3$rd $p$-wave resonance respectively. The magnitude of $A^{(-)}_g$ complies with the linear dimension of the ground super-Efimov state at resonance. The convergence of $A^{(-)}_g$ to approximately $-42.0\,l_\text{vdW}^2$ is reminiscent of the Efimov physics in which the three-body parameters becomes more universal for two-body potentials that can support more bound states \cite{Wang1}. Recent successful realization of ``quasi" $2$D Fermi gases \cite{Turlapov2010, Vale2011, Kohl2011PRL} 
opens up the prospect of experimental study of the super-Efimov physics in atomic gases. It will be worth examining how the super-Efimov physics would be affected by the strong confinement applied to produce the ``quasi" $2$D gases in future investigations.

\emph{Acknowledgments.---}
We thank S.~Moroz, Y.~Nishida, H.~Zhai, X.~Cui, Z.~Shi, S.~Tan, Y.~Castin, C.H.~Greene, J.P.~D'Incao and R.~C\^{o}t\'{e} for discussions. ZY acknowledges support from the INT program ``Universality in Few-Body Systems: Theoretical Challenges and New Directions" (INT 14-1), during which part of the work was carried out. This work is supported by Tsinghua University Initiative Scientific Research Program, NSFC under Grant No.~11104157, No.~11474179, and No.~11204152, No.~11004118, No.~11174176, No.~11204153, and NKBRSFC under Grant No. 2011CB921500.


\begin{thebibliography}{99}

\bibitem{efimov}V.~Efimov, Phys.~Lett.~B \textbf{33} 563 (1970); Yad.~Fiz.~\textbf{12}, 1080 (1970) [Sov.~J.~Nucl.~Phys.~\textbf{12}, 589 (1971)]; Nucl.~Phys.~A \textbf{210}, 157 (1973).

\bibitem{braaten}E.~Braaten, and H.-W.~Hammer, Phys.~Rep.~\textbf{428}, 259 (2006).

\bibitem{petrov}D. S.~Petrov, arXiv:1206.5752.

\bibitem{bloch}I.~Bloch, J.~Dalibard, and W.~Zwerger, Rev.~Mod.~Phys.~\textbf{80}, 885 (2008).

\bibitem{molmer}M.~Saffman, T. G.~Walker and K.~M\o lmer, Rev.~Mod.~Phys.~\textbf{82}, 2313 (2010).

\bibitem{dalibard}J.~Dalibard, F.~Gerbier, and G.~Juzeli\=unas, and P.~\"Ohberg, Rev.~Mod.~Phys.~\textbf{83}, 1523 (2011).

\bibitem{Chin}C.~Chin, R.~Grimm, P.~Julienne, and E.~Tiesinga,
Rev.~Mod.~Phys.~\textbf{82}, 1225 (2010).

\bibitem{grimm_cesium}T.~Kraemer, M.~Mark, P.~Waldburger, J.G.~Danzl, C.~Chin, B.~Engeser, A.D.~Lange, K.~Pilch, A.~Jaakkola, H.-C.~N\"agerl and R.~Grimm, Nature \textbf{440}, 315 (2006).

\bibitem{Jochim_3b}T. B.~Ottenstein, T.~Lompe, M.~Kohnen, A.N.~Wenz, and S.~Jochim, Phys.~Rev.~Lett.~\textbf{101}, 203202 (2008).

\bibitem{hulet}S. E.~Pollack, D.~Dries, and R.G.~Hulet, Science \textbf{326}, 1683 (2009).

\bibitem{Gross1} N.~Gross, Z.~Shotan, S.~Kokkelmans, and L.~Khaykovich, Phys.~Rev.~Lett.~\textbf{103}, 163202 (2009).
 
\bibitem{OHara1} J. H.~Huckans, J. R.~Williams, E. L.~Hazlett, R. W.~Stites, and K. M.~O'Hara, Phys.~Rev.~Lett.~\textbf{102}, 165302 (2009).
 
\bibitem{OHara2}J. R.~Williams, E. L.~Hazlett, J. H.~Huckans, R. W.~Stites, Y.~Zhang, and K. M.~O'Hara, Phys.~Rev.~Lett.~\textbf{103}, 130404 (2009).
 
\bibitem{Jin} R. J.~Wild, P.~Makotyn, J. M.~Pino, E. A.~Cornell, and D. S.~Jin, Phys.~Rev.~Lett.~\textbf{108}, 145305 (2012).

\bibitem{grimm_ad}S.~Knoop, F.~Ferlaino, M.~Mark, M.~Berninger, H.~Sch\"obel, H.-C.~N\"agerl, R.~Grimm, Nature Phys.~\textbf{5}, 227 (2009).

\bibitem{jochim_ad}T.~Lompe, T. B.~Ottenstein, F.~Serwane, K.~Viering, A. N.~Wenz, G.~Z\=urn, S.~Jochim, Phys.~Rev.~Lett.~\textbf{105}, 103201 (2010).

\bibitem{jochim_radio}T.~Lompe, T. B.~Ottenstein, F.~Serwane, A. N.~Wenz, G.~Z\=urn, S.~Jochim, Science \textbf{330}, 940 (2010).

\bibitem{ueda_radio}S.~Nakajima, M.~Horikoshi, T.~Mukaiyama, P.~Naidon, and M.~Ueda, Phys.~Rev.~Lett.~\textbf{106}, 143201 (2011).


\bibitem{gross}N.~Gross, Z.~Shotan, S.~Kokkelmans, and L.~Khaykovich, Phys.~Rev.~Lett.~\textbf{105}, 103203 (2010).

\bibitem{Berninger} M~Berninger, A~Zenesini, B.~Huang, W.~Harm, H.-C.~N\"agerl, F.~Ferlaino, R.~Grimm, P.S.~Julienne, and J. M. Hutson, Phys.~Rev.~Lett.~\textbf{107}, 120401 (2011). 

\bibitem{Wang1}J.~Wang, J. P.~D`Incao, B.D.~Esry, and C. H.~Greene, Phys.~Rev.~Lett.~\textbf{108}, 263001 (2012).

\bibitem{Wang2}Y.~Wang, J.~Wang, J. P.~D`Incao, and C.H.~Greene, Phys.~Rev.~Lett.~\textbf{109}, 243201 (2012).

\bibitem{Chin_a}C.~Chin, arXiv:1111.1484.

\bibitem{Schmidt}R.~Schmidt, S. P.~Rath, and W.~Zwerger, Eur.~Phys.~J.~B \textbf{85},386 (2012).

\bibitem{Naidon}P.~Naidon, S.~Endo, and M.~Ueda,  Phys.~Rev.~Lett.~\textbf{112}, 105301 (2014).

\bibitem{nishida_super} Y.~Nishida, S.~Moroz, and D. T.~Son, Phys.~Rev.~Lett.~\textbf{110}, 235301 (2013).

\bibitem{frankfurt}  D. K. Gridnev, J.~Phys.~A \textbf{47}, 505204 (2014).

\bibitem{nielsen}E.~Nielsen, D. V.~Fedorov, A. S.~Jensen, and E.~Garrido, Phys.~Rep.~\textbf{347},  373 (2001).

\bibitem{zinner}  A.G. Volosniev, D. V. Fedorov, A. S. Jensen and N. T. Zinner, J. Phys. B \textbf{47}, 185302 (2014).

\bibitem{castin} Y.~Castin, private communication.

\bibitem{jesper} J.~Levinsen, N. R.~Cooper, and V.~Gurarie, Phys.~Rev.~A \textbf{78}, 063616 (2008).


\bibitem{Johnson} B. R.~Johnson, J.~Chem.~Phys.~\textbf{73}, 5051 (1980).

\bibitem{Lepetit} B.~Lepetit, Z.~Peng, A.~Kuppermann, Chem.~Phys.~Lett.~\textbf{166}, 572 (1990).

\bibitem{Lin} C. D.~Lin, Phys.~Rep.~\textbf{257}, 1 (1995).

\bibitem{Suno}H.~Suno and B. D.~Esry, Phys.~Rev.~A \textbf{78}, 062701 (2008).

\bibitem{Wang_Dincao}J.~Wang, J. P.~D'Incao and C. H.~Greene Phys. Rev. A \textbf{84}, 052721 (2011).

\bibitem{Dincao1}J. P.~D'Incao and B. D.~Esry, Phys. Rev. A \textbf{90}, 042707 (2014).

\bibitem{Dincao2}J. P.~D'Incao, F. Anis, and B. D.~Esry, arXiv: 1411.2321.

\bibitem{Wang_Wang}J.~Wang, J. P.~D'Incao, Y.~Wang and C. H.~Greene, Phys.~Rev.~A \textbf{86}, 062511 (2012).

\bibitem{Fano}U.~Fano, Phys.~Rev.~\textbf{124}, 1866 (1961).

\bibitem{Turlapov2010}K.~Martiyanov, V.~Makhalov, and A.~Turlapov, Phys.~Rev.~Lett.~\textbf{105}, 030404 (2010).

\bibitem{Vale2011}P.~Dyke, E. D.~Kuhnle, S.~Whitlock, H.~Hu, M.~Mark, S.~Hoinka, M.~Lingham, P.~Hannaford, C.J.~Vale, Phys.~Rev.~Lett.~\textbf{106}, 105304 (2011).

\bibitem{Kohl2011PRL}B. Fr\"ohlich, M. Feld, E. Vogt, M. Koschorreck, W. Zwerger, and M. K\"ohl, Phys. Rev. Lett. ~\textbf{106}, 105301 (2011).

\end{thebibliography}
\end{document}